\documentclass[12pt]{article}
\usepackage{epsfig}
 
\newlength{\dinwidth} \newlength{\dinmargin}
\begin{document}
 
\begin{center}
{\Large \bf Theoretical status of the top quark cross 
section}\footnote{Presented at DPF 2004, Riverside, California,
August 26-31, 2004.}
\end{center}
\vspace{2mm}

\begin{center}
{\large Nikolaos Kidonakis$^a$ and Ramona Vogt$^b$}\\
\vspace{2mm}
{\it 
$^a$ Kennesaw State University, 1000 Chastain Rd., \#1202\\
Kennesaw, GA 30144-5591, USA\\
and\\
Cavendish Laboratory, University of Cambridge\\
Madingley Road, Cambridge CB3 0HE, England\\
\vspace{2mm}
$^b$ 
Nuclear Science Division, Lawrence Berkeley National Laboratory,\\ 
Berkeley, CA 94720, USA \\
and \\
Physics Department, University of California at Davis,\\ 
Davis, CA 95616, USA
}
\end{center}
  
\vspace{3mm}
  
\begin{abstract}

We discuss the most recent calculations of the top quark total cross section 
and transverse momentum distributions at the Tevatron and the LHC. These 
calculations include the soft-gluon corrections at next-to-next-to-leading 
order (NNLO). The soft NNLO corrections stabilize the scale dependence of
the cross section.

\end{abstract}

\thispagestyle{empty} \newpage \setcounter{page}{2}

\section{Introduction}

Top quark pair production at the Tevatron and the LHC
are processes that will establish fundamental properties 
of the top quark, including the mass and the cross section.
The top quark is now actively studied in Run II at the Tevatron \cite{topexp}.
Top quark production at the Tevatron receives large corrections
from soft gluons in the near threshold region \cite{KS,NK}. 
The best estimates of the
cross section include these corrections beyond next-to-leading order.

The calculation has been performed in both 
single-particle-inclusive (1PI) and pair-invariant-mass (PIM) 
kinematics \cite{NKRV,NNNLL}.
In 1PI kinematics, 
$i(p_a) + j(p_b) \rightarrow t(p_1) + X[{\overline t}](p_2)$
with $ij=q{\bar q}$ or $gg$.
We define $s=(p_a+p_b)^2$, $t_1=(p_b-p_1)^2-m^2$, $u_1=(p_a-p_1)^2-m^2$
and $s_4=s+t_1+u_1$. At threshold, $s_4 \rightarrow 0$ and 
the  soft corrections take the form
$[\ln^l(s_4/m^2)/s_4]_+$.
In PIM kinematics,
$i(p_a) + j(p_b) \rightarrow t{\overline t}(M) + X(k)$.
At threshold, $z=M^2/s \rightarrow 1$ and
the soft corrections are
$[\ln^l(1-z)/(1-z)]_+$.
 
We denote the soft-gluon corrections by
${\cal D}_l(x_{\rm th})\equiv[\ln^l(x_{\rm th})/x_{\rm th}]_+$
where $x_{\rm th}$ is defined as $s_4$ or $1-z$, depending on the kinematics.
For the order $\alpha_s^n$ corrections, $l\le 2n-1$.
At NLO, we have ${\cal D}_1(x_{\rm th})$ and  ${\cal D}_0(x_{\rm th})$ terms.
At NNLO, we have leading ${\cal D}_3(x_{\rm th})$, 
next-to-leading ${\cal D}_2(x_{\rm th})$, next-to-next-to-leading
${\cal D}_1(x_{\rm th})$, and next-to-next-to-next-to-leading logarithms 
(NNNLL) ${\cal D}_0(x_{\rm th})$. At present, all NNLO soft logarithms
can be fully calculated except for some NNNLL two-loop \cite{NKtwoloop} 
process-dependent terms which are expected to be numerically small.
 
We can formally resum the soft logarithms to all orders in $\alpha_s$,
but a resummed result is prescription dependent. If we expand to fixed order,
however, we can derive prescription-independent results.
A unified approach and a master formula
for calculating these logarithms to NNLO in the fixed-order expansion 
for any process was presented in Ref. \cite{NKuni}. 
It was applied to top pair production in Ref. \cite{NNNLL}
where logarithms through NNNLL at NNLO were calculated along 
with some virtual terms. We call this a NNLO-NNNLL+$\zeta$ calculation.
Similar studies of related heavy quark and electroweak processes, 
including bottom and charm production \cite{botcharm}, FCNC top quark 
production \cite{NKAB}, charged Higgs production with a top 
quark \cite{cHiggs}, and electroweak-boson production \cite{NKOSV}, 
have also recently been completed.

\section{Top quark pair production cross section}
 
\begin{figure}
\centerline{\psfig{file=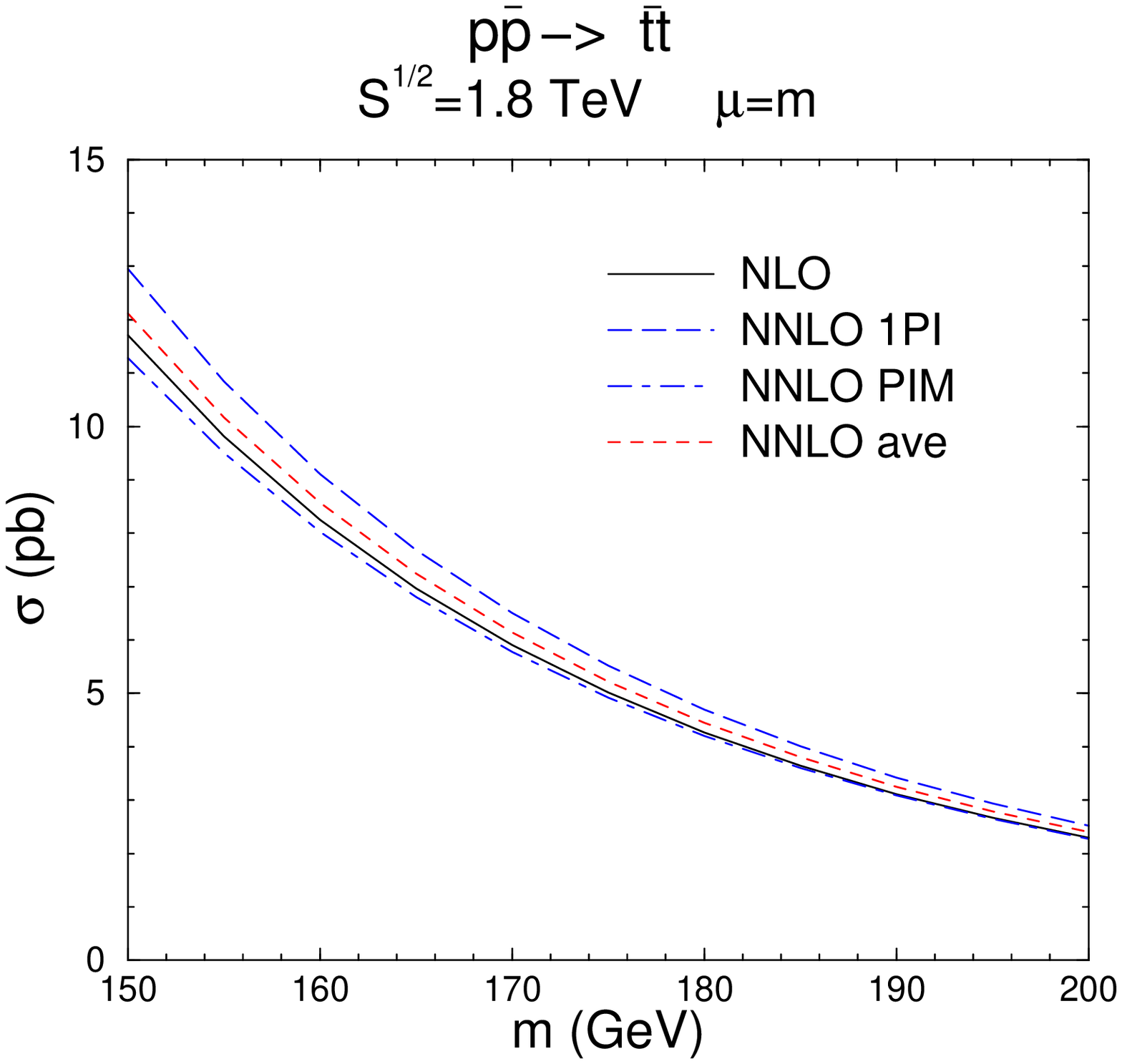,width=8cm} \hspace{10mm}
\psfig{file=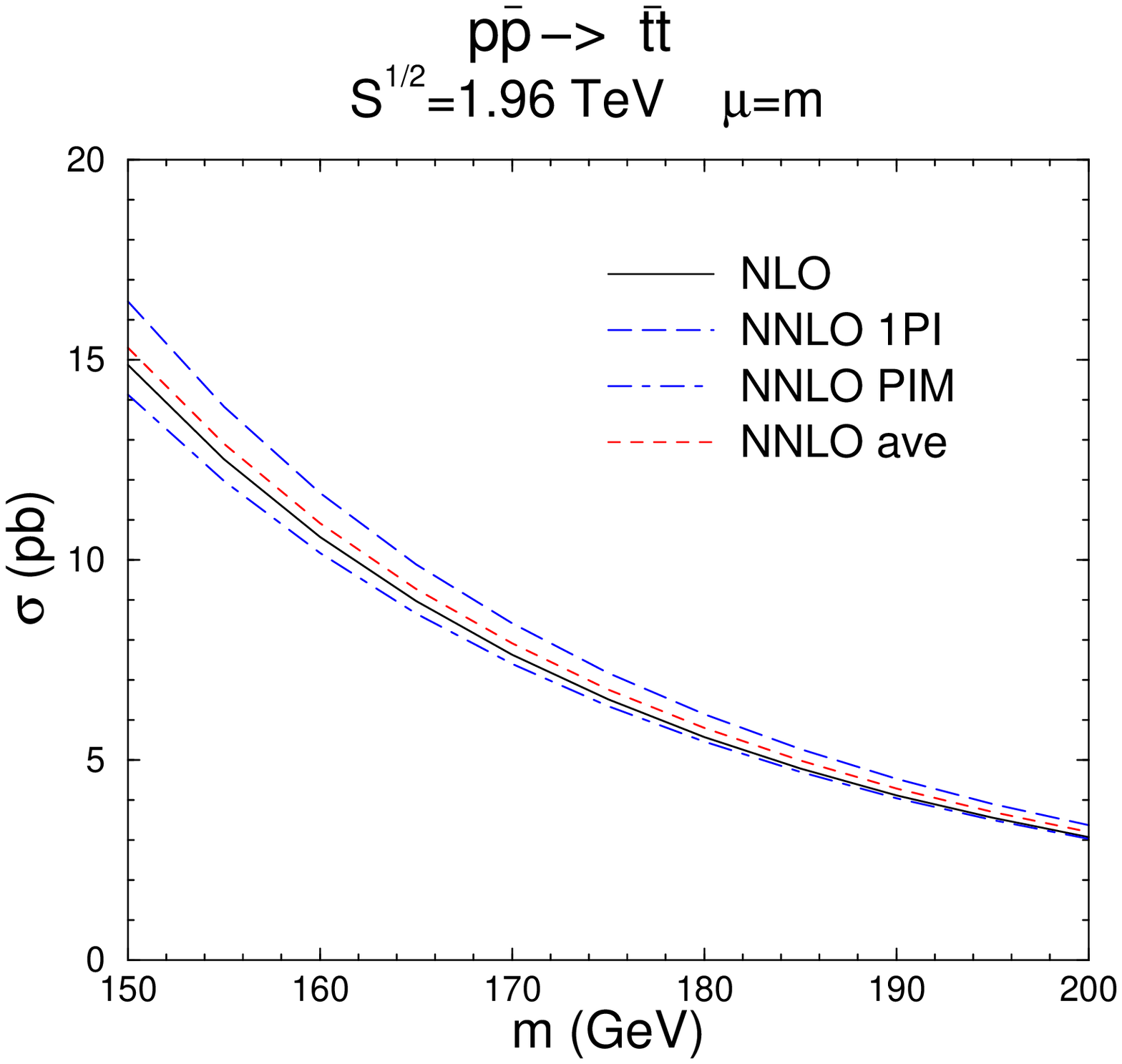,width=8cm}}
\vspace*{5pt}
\caption{Top quark pair production cross section at the Tevatron to NLO
and NNLO.  Here
``NNLO ave'' denotes the average of the 1PI and PIM results at NNLO.  The
left-hand side shows $\sqrt{S} = 1.8$ TeV while $\sqrt{S} = 1.96$ TeV is 
shown on the right-hand side.  The results are shown for $\mu = m$.}
\end{figure}

We now present our results for
$p {\bar p} \rightarrow t{\bar t}$ production at the Tevatron.
In Fig. 1 we plot the pair cross section as a function of top mass
at 1.8 TeV (left) and 1.96 TeV (right).
For $m = 175$ GeV, we find \cite{NNNLL} 
$\sigma(\sqrt{S}=1.8$ TeV$)= 5.24 \pm 0.31$ pb and
$\sigma(\sqrt{S}=1.96$ TeV$)=6.77 \pm 0.42$ pb.  The
uncertainty is the kinematics ambiguity.

 \begin{figure}
\centerline{\psfig{file=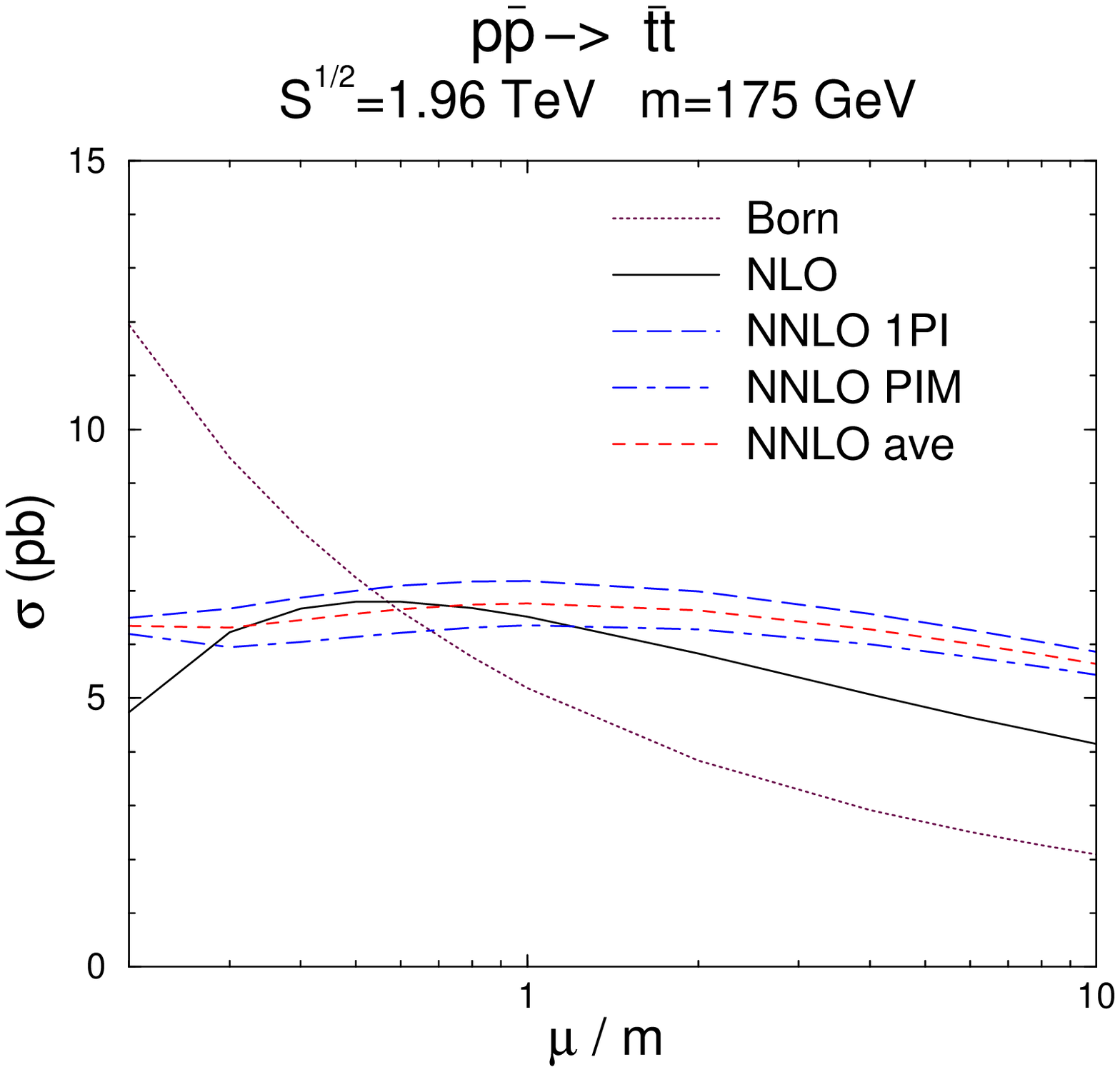,width=8cm} \hspace{10mm}
\psfig{file=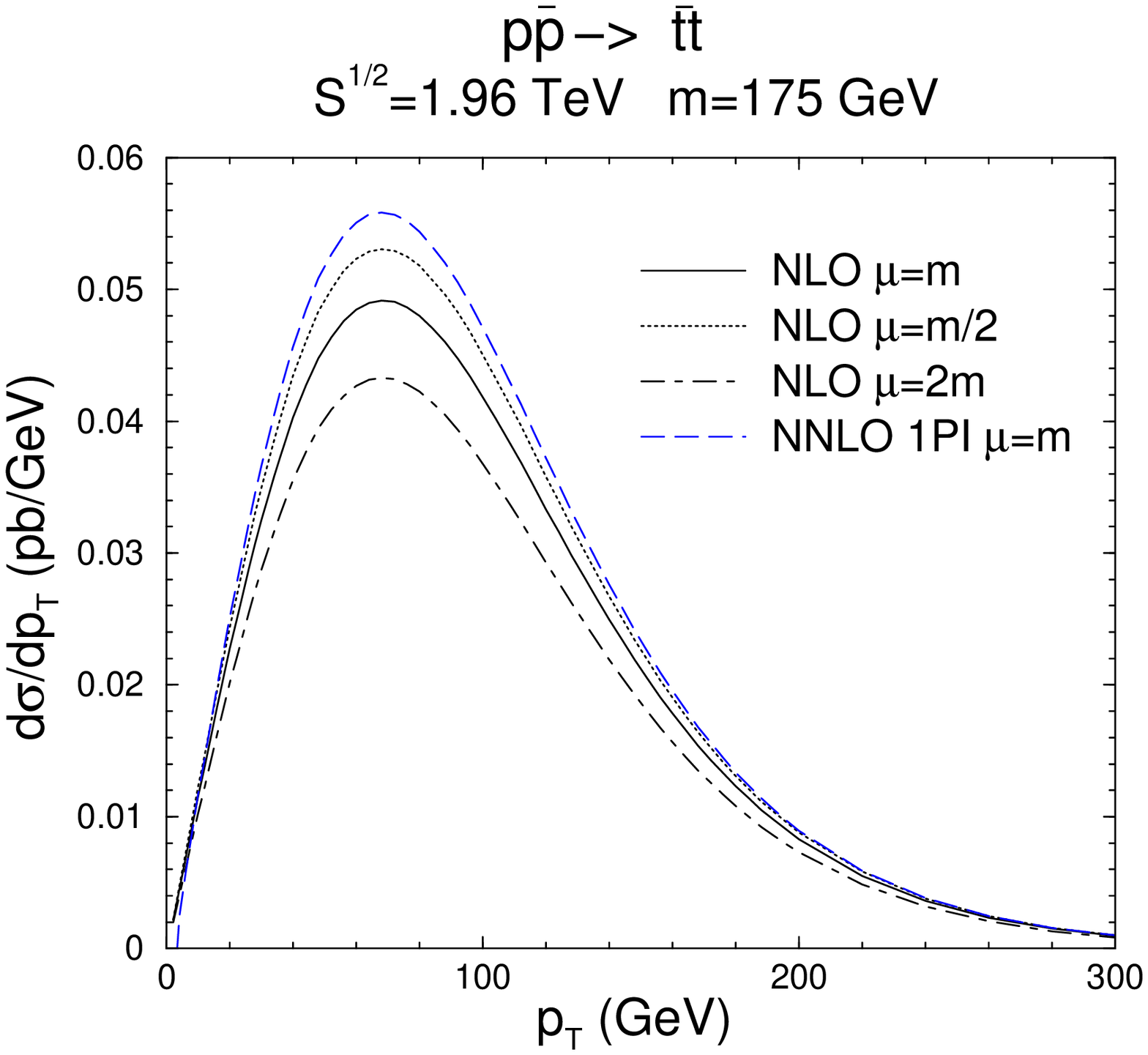,width=8cm}}
\vspace*{5pt}
\caption{The scale dependence of the $t \overline t$ cross section (left) and 
the top quark $p_T$ distribution in 1PI kinematics (right)
at $\sqrt{S} = 1.96$ TeV.}
\end{figure} 

The scale dependence at the Tevatron Run II energy is shown on the left-hand 
side of Fig. 2.  It is clear that the NNLO result is much more stable than 
the NLO.  The top quark tranverse momentum, $p_T$, distribution
at $\sqrt{S} = 1.96$ TeV, calculated in 1PI kinematics, 
is shown on the right-hand side of Fig. 2.
 
Finally, our best estimate for the $t \overline t$ total cross section 
at the LHC, $\sqrt{S} = 14$ TeV, is 873 pb for $m= 175$ GeV \cite{NNNLL}.


\begin{thebibliography}{0}
 
\bibitem{topexp}
E. Shabalina, Eur. Phys. J. {\bf C33}, s472 (2004); \newline
P. Azzi, Int. J. Mod. Phys. {\bf A19}, 785 (2004); \newline
D. Gerdes, FERMILAB-CONF-04-011-E; \newline
J. Nielsen, FERMILAB-CONF-04-068-E; \newline
J. Thom, FERMILAB-CONF-04-089-E; \newline
CDF Coll., hep-ex/0404036; \newline
J. Ellison, hep-ex/0405077; \newline
I. Iashvili, hep-ex/0407056; \newline
A. Hocker, hep-ex/0408151; \newline
P. Movilla Fernandez, hep-ex/0409001; \newline
L. Sonnenschein, hep-ex/0410036. 

\bibitem{KS}
N. Kidonakis and G. Sterman, {\it Phys. Lett.} {\bf B387}, 867 (1996);
{\it Nucl. Phys.} {\bf B505}, 321 (1997);
N. Kidonakis, G. Oderda, and G. Sterman,
{\it Nucl. Phys.} {\bf B531}, 365 (1998).
 
\bibitem{NK}
N. Kidonakis, {\it Int. J. Mod. Phys.} {\bf A15}, 1245 (2000);
{\it Mod. Phys. Lett.} {\bf A19}, 405 (2004);
in DPF 2004, hep-ph/0410116.
 
\bibitem{NKRV}
N. Kidonakis, {\it Phys. Rev.} {\bf D64}, 014009 (2001);
{\it Int. J. Mod. Phys.} {\bf A16} Suppl. 1A, 363 (2001);
N. Kidonakis, E. Laenen, S. Moch, and R. Vogt,
{\it Phys. Rev.} {\bf D64}, 114001 (2001);
{\it Phys. Rev.} {\bf D67}, 074037 (2003);
{\it Nucl. Phys.} {\bf A715}, 549c (2003).
 
\bibitem{NNNLL}
N. Kidonakis and R. Vogt, {\it Phys. Rev.} {\bf D68}, 114014 (2003);
{\it Eur. Phys. J.} {\bf C33}, s466 (2004).

\bibitem{NKtwoloop}
N. Kidonakis, hep-ph/0208056; in DIS 2003, hep-ph/0307145.

\bibitem{NKuni}
N. Kidonakis, {\it Int. J. Mod. Phys.} {\bf A19}, 1793 (2004);
in DIS 2003, hep-ph/0306125, hep-ph/0307207.
 
\bibitem{botcharm}
N. Kidonakis and R. Vogt, {\it Eur. Phys. J.} {\bf C36}, 201 (2004);
in DIS 2004, hep-ph/0405212.

\bibitem{NKAB}
A. Belyaev and N. Kidonakis, {\it Phys. Rev.} {\bf D65}, 037501 (2002);
N. Kidonakis and A. Belyaev, {\it JHEP} {\bf 12}, 004 (2003);
in DIS 2004, hep-ph/0407032.

\bibitem{cHiggs}
Higgs Working Group, hep-ph/0406152;
N. Kidonakis, in DIS 2004, hep-ph/0406179.

\bibitem{NKOSV}
N. Kidonakis and A. Sabio Vera, {\it JHEP} {\bf 02}, 027 (2004);
in DIS 2004, hep-ph/0405013; in DPF 2004, hep-ph/0409206;
hep-ph/0409337; N. Kidonakis and J.F. Owens,
{\it Int. J. Mod. Phys.} {\bf A19}, 149 (2004).


\end{thebibliography}
\end{document}